\documentclass[12pt]{article}
\usepackage{graphicx}
\usepackage{epsfig}

\usepackage{amssymb}

\begin{document}

\title{On the scaling range of power-laws originated from fluctuation analysis}
\author{Dariusz Grech$^{(a)}$\footnote{dgrech@ift.uni.wroc.pl} and Zygmunt Mazur$^{(b)}$\footnote{zmazur@ifd.uni.wroc.pl}}
\date{}

\maketitle
\begin{center}
\begin{flushleft}
(a) Institute of Theoretical Physics, University of Wroc{\l}aw, Pl. M.Borna 9,\\ PL-50-204 Wroc{\l}aw, Poland

(b) Institute of Experimental Physics, University of Wroc{\l}aw, Pl. M.Borna 9, \\PL-50-204 Wroc{\l}aw, Poland
\end{flushleft}
\end{center}

\hspace{1.5 cm}

\begin{abstract}
We extend our previous study of scaling range properties  done for detrended fluctuation analysis (DFA) \cite{former_paper} to other techniques of fluctuation analysis (FA). The new technique called Modified Detrended Moving Average Analysis (MDMA) is  introduced and its scaling range properties are examined and compared with those of detrended moving average analysis (DMA) and DFA. It is shown that contrary to DFA, DMA and MDMA techniques exhibit power law dependence of the scaling range with respect to the length of the searched signal and with respect to the accuracy $R^2$ of the fit to the considered scaling law imposed by DMA or MDMA schemes. This power law dependence is satisfied for both uncorrelated and autocorrelated data. We find also a simple generalization of this power law relation for series with different level of autocorrelations measured in terms of the Hurst exponent. Basic relations between scaling ranges for different techniques are also discussed. Our findings should be particularly useful for local FA in e.g., econophysics, finances or physiology, where the huge number of short time series has to be examined at once and wherever the preliminary check of the scaling range regime for each of the series separately is neither effective nor possible.
\end{abstract}
$$
$$


\section{Introduction and motivation.}

The presence of long memory effects in complex systems has been studied by many authors in variety of contexts and in various areas of science. Complexity of the system can be translated into properties of data series exhausted from the system. Therefore, a lot of information on complexity can be drawn from the precise study of memory effects in data. The latter effects are most easily searched in stationary time series $x_i$, $(i=1,...,L)$, by two point autocorrelation function $C_s=\langle\Delta x_i \Delta x_{i+s}\rangle$, where $\langle\, \rangle$ is the average taken from all data in the signal with series increments $\Delta x_i = x_{i+1}-x_i$. The autocorrelation function changes with time lag $s$ according to the power law \cite{gamma_1,gamma_2}:
\begin{equation}
C_s\simeq (2-\gamma)(1-\gamma)s^{-\gamma},
\end{equation}
where the autocorrelation scaling exponent $\gamma$ $(0\leq \gamma \leq 1)$ stands for the level of memory in signal and its two edge values correspond respectively to fully correlated  $(\gamma =0)$ or completely uncorrelated $(\gamma=1)$ data.

The estimation of scaling exponent $\gamma$ can be done using various methods.  The most popular are based on the so-called fluctuation analysis (FA) \cite{FA}.
In the FA approach the sequence $\Delta x_i$ is treated as steps of a discrete random walk, i.e., one builds the cumulated series $x_t=\sum_{i=1}^t \Delta x_i$ $(t=1,...,L)$; then the variance of its displacement is
found by averaging over different time windows of length $t$.
The scaling exponent $H$ of the series, called also the
Hurst exponent \cite{Hurst1,Hurst2} can be measured because of power law relation:
\begin{equation}
var(x_t) = \langle x_t^2\rangle-\langle x_t\rangle^2\simeq t^{2H},
\end{equation}
where $var(·)$ is the variance. The scaling exponent $H$ is related to the autocorrelation exponent $\gamma$ by the linear relationship \cite{rel_gamma_H}:
\begin{equation}
H=1-\frac{\gamma}{2}
\end{equation}

One has to keep in mind that a direct calculation of correlation functions and $\gamma$ exponent is hindered by the
level of noise present in time series and
by possible non-stationarities in the data. To
reduce these effects one usually does not calculate $\gamma$ directly, but
 study instead the "integrated profile" of the data, i.e. $x_t$ instead of $\Delta x_i$ series. This approach makes the fundament of all FA methods.

One should also remember that time series $x_t$ has to be detrended before using Eq.(2), otherwise an artificial bias is introduced leading to incorrect values of the scaling exponents. The so-called detrendization procedure, i.e. subtraction of trend in given data, was first proposed in detrended fluctuation analysis (DFA)\cite{DFA_1,DFA_2,DFA_3} and then in detrended moving average analysis (DMA)\cite{DMA_1,DMA_2,DMA_3,DMA_4}. In the case of DFA, the trend is defined as the polynomial fit\footnote{usually the linear fit is sufficient} to data in the considered time window of length $\tau$.
The mean-square fluctuation $F^2(\tau)$ of the signal around its trend (detrended signal) is calculated and then $F^2(\tau)$ is averaged over all time windows of length $\tau$.

One expects, according to Eq.(2), that the power law
\begin{equation}
\langle F_{DFA}^2(\tau)\rangle_{box}\sim \tau^{2H}
\end{equation}
is fulfilled, where $\langle.\rangle_{box}$ is the expectation value, i.e. the average taken over all time windows of fixed size. The above equation provides one of the most frequently used methods in stationary and even nonstationary series to calculate their characteristic main $H$ exponent and the autocorrelation properties induced by Eq.(1).

It was examined in our former article \cite{former_paper} that in the case of DFA the scaling range (i.e. the range of $\tau$ values satisfying Eq.(4)) is a linear function of data length $L$ and the goodness of line fit $R^2$ in log-log scale to Eq.(4). More precisely, the formula
\begin{equation}
 \lambda_{DFA}(u,L,H) = ((\alpha H + \beta)u+ \alpha_0)L + \gamma
\end{equation}
with $u=1-R^2$ was found for DFA scaling range ($\lambda_{DFA}$) and the coefficients $\alpha$, $\beta$, $\alpha_0$, $\gamma$ were estimated numerically (see Ref.\cite{former_paper} for details).

In DMA, contrary to DFA, a trend is found as the moving average of the assumed length $\lambda$ and is calculated from data points immediately preceding the given one, say $x_i$. It means that statistics of data points entering the calculation of detrended signal  depends strongly on the chosen length $\lambda$ of the moving average. For given $\lambda$ only data points $x_i$ with $i\geq\lambda$ can be taken into account for detrended fluctuation  according to \cite{DMA_1}

\begin{equation}
F_{DMA}^2(\lambda) =\frac{1}{L-\lambda +1}{\sum_{i=\lambda}^L (x_i - \langle x_i\rangle_{\lambda})^2}
\end{equation}
where $\langle x_i\rangle_{\lambda}$ is the moving average of length $\lambda$ defined as
\begin{equation}
\langle x_i\rangle_{\lambda}=\frac{1}{\lambda}\sum_{k=i-\lambda +1}^{i} x_k
\end{equation}
and plays the role of a trend in data series.

The power law  similar to DFA is expected \cite{DMA_1, DMA_2}:
\begin{equation}
F_{DMA}^2(\lambda)\sim \lambda^{2H}
\end{equation}

The limitations in statistics imposed by DMA can easily be skipped in a simple and straightforward way, quite acceptable in practical applications. Such modification of DMA, called by us \textit{modified} DMA (MDMA) is presented in the next section. Then we analyze the scaling range properties of these two methods and compare them with DFA results. Similarly to our previous approach \cite{former_paper}, the most interesting case to examine is the scaling range property for short series what is particularly useful for local FA,  e.g. in econophysics and finances \cite{finance_1, DFA_loc2, grech_1, grech_2, DFA_loc4, grech_3, kristoufek} or physiology \cite{heart_1,heart_2}. Often a huge amount of short time series has to be examined at once in these applications and the preliminary check of the scaling range regime for each of the series separately is neither possible nor effective.

\section{Introducing modified detrended moving average \\analysis (MDMA)}

DMA has a diversified statistics for detrended data entering  Eq.(6) with various lengths $\lambda$ of moving averages. For the particular choice of $\lambda$ and for the series length $L\geq \lambda$ only $L-\lambda +1$ detrended data points enter the mean-square fluctuation $F_{DMA}^2$ in Eq.(6). Hence, the more reliably a trend is determined within DMA, the weaker statistics is available for estimation of the scaling exponent $H$. Obviously, it is very uncomfortable situation. Especially, it leads to shorter scaling regimes for discussed scaling power laws for larger $\lambda$ values\footnote{scaling properties for small $\lambda$ are also inappropriate because moving averages with small $\lambda$ do not reconstruct real trends}. This obstacle may be rather easy eliminated in the new proposal explained below.

 One usually deals in practise with time series where, for various reasons, only a part of data is taken for further processing. In finance for instance, we investigate series of data starting at some moment $t_0$ and terminating, say, at $t_0+\Delta t$. Often, it does not mean that earlier data for $t<t_0$ are not available. We do not explore  them but they usually exist and can be used as the background to calculate the necessary trends (moving averages) if one wants to. Therefore, we propose to  modify slightly the scheme calculating $F_{DMA}^2$ in Eq.(6). The new version called MDMA assumes that some amount of data is stored before the basic time series $\{x_i\}$ $(i=1,...,L)$. The whole available amount of data can therefore be written as the sequence $\{x_{-\lambda_{max}},...,x_{-2},x_{-1},x_1,x_2,...,x_L\}$, where $\lambda_{max}$ is the maximal used scaling range to be determined in this article. Such an approach enables to calculate trend (moving averages) for those data points where DMA procedure with particular choice of $\lambda$ simply fails. To be precise, Eq.(6) is replaced within MDMA by
\begin{equation}
F_{MDMA}^2(\lambda) =\frac{1}{L}\sum_{i=1}^L (x_i - \widetilde{\langle x_i\rangle}_{\lambda})^2
\end{equation}
where the moving average of length $\lambda$ is modified to:
\begin{equation}
\widetilde{\langle x_i\rangle}_{\lambda}=\frac{1}{\lambda}\sum_{k=i-\lambda +1, k\neq0}^{i}x_k
\end{equation}
and $k<0$ indicates the sum running over additional data preceding the basic series.

The power law in Eq.(8) is still kept when $F_{MDMA}^2(\lambda)$ is substituted for $F_{DMA}^2(\lambda)$:
\begin{equation}
F_{MDMA}^2(\lambda)\sim \lambda^{2H}
\end{equation}

It is easily seen just from the construction recipe that MDMA produces the same scaling exponents as DMA but with bigger accuracy due to larger scaling range available for calculations. We will refer to this issue in the following sections.

\section{DMA and MDMA scaling ranges for uncorrelated and autocorrelated data}

We begin with statistical analysis of an ensemble of artificially generated uncorrelated time series with the given length $L$. Then we find in the same way as in Ref.\,\cite{former_paper} the percentage rate of series which are below the specified level of regression line fit $R^2$ in log-log scale,  induced by Eq.(8) in the case of DMA and by Eq.(9) in the case of MDMA, assuming that the maximal length of moving average $\lambda_{max}$ is being fixed. Fig.\,1 illustrates the rejection rate, i.e. the percentage rate of series not matching the assumed criterion for $R^2$, calculated for the  series of uncorrelated $(H=1/2)$ data of length $L=10^3$ for different $R^2$ values and for running  $\lambda_{max}$. DMA and MDMA cases are shown for comparison to indicate that the scaling range for MDMA is bigger than for DMA, assuming the same requirements of the goodness fit $R^2$.

In further analysis we took two specific rejection thresholds: $2.5\%$ and $5\%$ corresponding to confidence levels (CL): $97.5\%$ and $95\%$  respectively. All data have been gathered from the ensemble of  $5\times 10^4$  generated time series with a length varying between $L=5\times 10^2$ and $L=10^4$.

Let us introduce new parameter $u=1-R^2$ and investigate the $\lambda(u,L)$ dependence for DMA and MDMA methods in the same way as we did before for DFA in Ref.\,\cite{former_paper}.
 We have made throughout this paper the lower limit restriction for the minimal length of used moving averages ($\lambda_{min}=8$) because below this threshold a
significant lack of scaling in DMA is observed \footnote{ shorter moving averages do not determine the local trend precisely what causes the appearance of  artificial autocorrelations in detrended data and makes the similar effect as too small time window boxes in  DFA}.
All values were taken from the analysis of dependencies like in Fig.\,1. The functional dependence  $\lambda(u,L)$ is not linear contrary to previous findings for DFA. Therefore we used log-log plots first to verify if it has power law origin. This hypothesis turned out to be very fruitful. Figs.\,2a,\,3a convince that the scaling range  $\lambda(u,L)$ for DMA and MDMA methods factorize as

\begin{equation}
 \log(\lambda(u,L)) = A(u)\log L + B(u)
\end{equation}
where, according to Figs.\,2b,\,3b, $A(u)$ and $B(u)$ are also linear in $\log(u)$.

Hence, we expect the precise form of Eq.(12):

\begin{equation}
 \log(\lambda(u,L)) = (a\log u + a_0)\log L + b\log u + b_0
\end{equation}
with some real constants $a$, $a_0$, $b$, $b_0$.

The more detailed exploration of $A(u)$ and $B(u)$ dependencies (see Fig.\,4a,b) provides arguments for just three free parameters entering the fit of Eq.(13), since $a=0$ for both considered confidence levels. Therefore, one arrives with the final power law formula linking  the scaling range $\lambda(u,L)$ of uncorrelated time series with $u$ and $L$ for DMA and MDMA:
\begin{equation}
\lambda(u,L)) = D L^\eta  u^\xi
\end{equation}
where $D$, $\eta$ and $\xi$ are constants related in obvious manner with $a_0$, $b$ and $b_0$.

We have checked that the same qualitative scenario is fulfilled also for long range correlated series ($H>1/2$). Our examination was limited to series with $0.5<H<0.9$ which are most often met in practise in variety of areas. Figs.\,5,\,6 show the extension of findings from Figs.\,2,\,3 to cases of signals with long memory. The  exemplary plots revealing the $A(u)$ and $B(u)$ relations for the autocorrelated series (the case $H=0.8$) are shown in Figs.\,4c,\,4d. An interesting challenge is to find the quantitative form of $D(H)$, $\eta(H)$ and $\xi(H)$ functions which generalize Eq.(14) when applied to series with memory. This task is shifted to section 4.

\begin{table}
\centering
\begin{tabular}{||c||c|c|c|c|c||c|c|c|c|c||}
   \hline
   $H\setminus CL$ & $D^{97.5\%}$ & $\eta^{97.5\%}$ & $\xi^{97.5\%}$ & $\Delta^{97.5\%}_{MAE}$ & $\Delta^{97.5\%}_{ME}$ & $D^{95\%}$ & $\eta^{95\%}$ & $\xi^{95\%}$ & $\Delta^{95\%}_{MAE}$ & $\Delta^{95\%}_{ME}$\\
\hline
$H=0.5$ & 0.879 & 1.062 & 0.723 & 1.4\% & 4.2\% & 0.961  & 1.064  & 0.680 & 1.6\% & 3.5\% \\
\hline
$H=0.6$ & 0.940 & 1.050 & 0.652 & 1.3\% & 3.5\% & 1.078 & 1.048 & 0.616 & 1.3\% & 3.2\%\\
\hline
$H=0.7$ & 1.077 & 1.035 & 0.586 & 1.6\% & 3.9\% & 1.189 & 1.031 & 0.556 & 1.4\% & 3.1\%\\
\hline
$H=0.8$ & 1.068 & 1.022 & 0.526 & 1.3\% & 3.5\%  & 1.299 & 1.015 & 0.502 & 1.3\% & 3.0\%\\
\hline
\end{tabular}
\caption{Results of the best fit for coefficients in Eq.(14) describing scaling range for DMA. The fit is done for series with various autocorrelation level measured by $H$ exponent and for chosen two confidence levels $CL$: $97.5\%$ and $95\%$. }
\label{tab1}
\end{table}

\begin{table}
\centering
\begin{tabular}{||c||c|c|c|c|c||c|c|c|c|c||}
   \hline
   $H\setminus CL$ & $D^{97.5\%}$ & $\eta^{97.5\%}$ & $\xi^{97.5\%}$ & $\Delta^{97.5\%}_{MAE}$ & $\Delta^{97.5\%}_{ME}$ & $D^{95\%}$ & $\eta^{95\%}$ & $\xi^{95\%}$ & $\Delta^{95\%}_{MAE}$ & $\Delta^{95\%}_{ME}$\\
\hline
$H=0.5$ & 1.924 & 1.052 & 0.866 & 1.7\% & 4.6\% & 2.656  & 1.053  & 0.869 & 1.4\% & 4.0\%\\
\hline
$H=0.6$ & 2.310 & 1.039 & 0.808 & 1.5\% & 4.1\% & 3.131 & 1.037 & 0.805 & 1.1\% & 3.0\%\\
\hline
$H=0.7$ & 2.719 & 1.026 & 0.751 & 1.4\% & 3.4\% & 3.675 & 1.024 & 0.749 & 1.3\% & 3.6\%\\
\hline
$H=0.8$ & 3.086 & 1.012 & 0.694 & 1.1\% & 2.8\%  & 4.203 & 1.013 & 0.699 & 1.4\% & 3.2\%\\
\hline
\end{tabular}
\caption{Results of the best fit for coefficients in Eq.(14) describing scaling range for MDMA. Same notation applies as in Table 1.}
\label{tab1}
\end{table}

The best fit parameters to Eq.(14)  are gathered in Table 1 for DMA  and in Table 2 for MDMA  for two distinct confidence levels.
We required simultaneously minimization of the mean relative absolute error (MAE) and the maximal relative error (ME) for each of the fitting points in the same way as it was formerly done by us for DFA (see \cite{former_paper} for details). The fit was based on nearly $100$ data pairs $(u_i,L_j)$, where $u_i=5\times 10^{-3}(1+i)$ with $i=1,2,...,9$ and  $L_j=500,600,800,1000,1200,1500,1800,2400,3000,6000$,$10000$ respectively.

The MAE denoted as $\Delta_{MAE}(\lambda)$ was specified as
\begin{equation}
\Delta_{MAE}(\lambda)=1/N_{(ij)} \sum_{ij}|(\lambda^{exp}_{ij}(u,L)-\lambda_{ij}(u,L))/\lambda_{ij}(u,L)|
\end{equation}
where $\lambda_{ij}(u,L)\equiv \lambda(u_i,L_j)$ is the fitting value of Eq.(12) for particular $u_i$ and $L_j$, $N_{(ij)}$ counts different $(ij)$ pairs and $\lambda^{exp}_{ij}(u,L)$ is the respective value of scaling range simulated numerically from an ensemble of time series.

Similarly, ME marked as $\Delta_{ME}$ was defined as
\begin{equation}
\Delta_{ME} = \max_{(ij)}(|\lambda^{exp}_{ij}(u,L)-\lambda_{ij}(u,L))|/\lambda_{ij}(u,L))
\end{equation}
In the case of autocorrelated signal all data were generated by Fourier filtering method (FFM) algorithm \cite{ffm}.

\section{Towards unified model of scaling ranges}

 Finally, we will look for an unified formula containing a minimal number of free parameters  and describing  scaling ranges of both uncorrelated and autocorrelated data within DMA or MDMA techniques.

 All data taken from Tables 1-2, when drawn against the Hurst exponent values (see Fig.7), prove the existence of simple linear relations
 \begin{equation}
 D^{(m)}=D_0^{(m)} + D^{(m)}_1 H
 \end{equation}

 \begin{equation}
 \eta^{(m)}=\eta^{(m)}_0 - \eta^{(m)}_1 H
 \end{equation}

 \begin{equation}
 \xi^{(m)}=\xi^{(m)}_0 - \xi^{(m)}_1 H
 \end{equation}
where $m=1, 2$ corresponds respectively to DMA and MDMA methods .

Note that $D^{(2)}_0=0$, while $D^{(1)}_0 \neq 0$ for all confidence levels. This simplifies the unified formula for scaling ranges in MDMA method
\begin{equation}
\lambda_{MDMA}(u,L)=D^{(2)}_1 H L^{\eta^{(2)}_0 - \eta^{(2)}_1 H}u^{\xi^{(2)}_0 - \xi^{(2)}_1 H}
\end{equation}
It contains only  $5$ free parameters, once the respective unified model for scaling ranges in DMA  contains one parameter  more ($D^{(1)}_0$):

\begin{equation}
\lambda_{DMA}(u,L)= (D^{(1)}_0 + D^{(1)}_1 H ) L^{\eta^{(1)}_0 - \eta^{(1)}_1 H}u^{\xi^{(1)}_0 - \xi^{(1)}_1 H }
\end{equation}

The values of these parameters, taken directly from the regression line fit of Figs.\,7a-7f, define the global fit to Eqs.(20)(21). They are specified in Table 3 for DMA and in Table 4 for MDMA, together with corresponding MAE and ME values.

\begin{table}
\centering
\begin{tabular}{||c||c|c|c|c|c|c|c|c||}
   \hline
   $CL$ & $D^{(1)}_0$ & $D^{(1)}_1$ & $\eta^{(1)}_0$ & $\eta^{(1)}_1$ & $\xi^{(1)}_0$ & $\xi^{(1)}_1$ & $\Delta^{(1)}_{MAE}$ & $\Delta^{(1)}_{ME}$\\
\hline
$97.5\%$ & 0.558 & 0.640 & 1.130 & 0.135 & 1.049 & 0.657  &  1.9\% & 5.8\% \\
\hline
$95\%$ & 0.400 & 1.125 & 1.146 & 0.163 & 0.975 & 0.593 & 1.6\% & 4.7\% \\
\hline
\end{tabular}
\caption{ Results of the best fit for coefficients in unified formula in Eq.(21) for DMA done for all data coming from series with various autocorrelation levels and for two chosen  confidence levels $CL$: $97.5\%$ and $95\%$.}
\label{tab3}
\end{table}

\begin{table}
\centering
\begin{tabular}{||c||c|c|c|c|c|c|c|c||}
   \hline
   $CL$ & $D^{(2)}_0$ & $D^{(2)}_1$ & $\eta^{(2)}_0$ & $\eta^{(2)}_1$ & $\xi^{(2)}_0$ & $\xi^{(2)}_1$ & $\Delta^{(2)}_{MAE}$ & $\Delta^{(2)}_{ME}$\\
\hline
$97.5\%$ & 0 & 3.847 & 1.118 & 0.133 & 1.148 & 0.566  &  1.6\% & 5.0\% \\
\hline
$95\%$ & 0 & 5.254 & 1.118 & 0.133 & 1.148 & 0.566 & 1.2\% & 4.3\% \\
\hline
\end{tabular}
\caption{Results of the best fit for coefficients in Eq.(20) describing scaling ranges for MDMA.}
\label{tab4}
\end{table}

 Let us notice that the fitting parameters for $\lambda_{MDMA}$  except $D^{(2)}_1$ coincide for both confidence levels.  This simplifies scaling range calculations for the newly introduced scheme.

\section{Final remarks, discussion and conclusions.}

In this study we searched for the scaling range properties of DMA technique which links fluctuations of detrended random walk $F^2(\lambda)$ with the length of the moving average $\lambda$ taken to subtract the trend in fluctuations. Our simulations have been made on the ensemble of $5\times 10^4$ short and medium-length time series of length $5\times 10^2\leq L \leq 10^4$ with varying level of long memory imposed  by Hurst exponent $0.5\leq H \leq 0.8$. The latter specific spread for $H$ values was used to refer to cases of long range autocorrelated data one may most often encounter in practise. We introduced also slightly modified version of DMA, called by us MDMA, and examined their scaling properties in comparison with DMA.

It turns out that scaling ranges for DMA reveal a simple power law relationship with the length of data $L$ and the goodness of linear fit $R^2$ for the fundamental equation (Eq.(8)) which links the Hurst exponent $H$ of time series with detrended fluctuations of a signal $F^2(\lambda)$ around its local trend. We found also that the similar relation is fulfilled for the newly proposed in this paper fluctuation technique called MDMA. A numerical fit to parameters describing the power law for scaling ranges in both methods was done.

It is evident from Eqs.(20)(21) and Tables 3-4 that the scaling ranges $\lambda_{DMA}$ of DMA and $\lambda_{MDMA}$ of MDMA satisfy $\lambda_{DMA}<\lambda_{MDMA}$ for  the same fixed length of data and the same goodness of fit (see Figs.\,8,\,9). More precisely, $\lambda_{MDMA}$ is around $10\%-20\%$ larger than $\lambda_{DMA}$ for uncorrelated data and $40\%-50\%$ larger than $\lambda_{DMA}$ for highly autocorrelated data.

It is also remarkable that the typical scaling range in DMA for $R^2 \sim 0.98$ is surprisingly short and does not exceed $10\%$ of the series length. It grows up to $\sim 20\%$ of series length if we loosen the requirements for goodness of fit to Eq.(8) down to $R^2 \sim 0.95$.

One may ask about a relationship between scaling ranges of DMA, MDMA and DFA -- the latter one considered in \cite{former_paper}. The summarized relationships between $\lambda_{DMA}$, $\lambda_{MDMA}$ and $\lambda_{DFA}$ based on Eqs.\,(5)(20)(21) are specified in Figs.\,8,\,9.
We notice that the mutual hierarchy between $\lambda_{DFA}$,  $\lambda_{DMA}$ and  $\lambda_{MDMA}$ is ambiguous  and strongly depends on the length of considered data $L$, the assumed goodness of fit $R^2$ (or $u$ parameter) and the level of autocorrelations in a signal.  We see that $\lambda_{DFA}$ dominates over $\lambda_{DMA}$ and $\lambda_{MDMA}$ for longer ($L\gtrsim 3\times 10^3$) uncorrelated series of data, while for strongly autocorrelated signals one obtains $\lambda_{DFA}<\lambda_{DMA}<\lambda_{MDMA}$ independently on the data length.

Finally we considered contours in $(u,L)$ plane representing solutions of the equality\\ $\lambda_{DMA}(u,L,H)=\lambda_{DFA}(u,L,H)$ for fixed values of $H$ exponent. The similar analysis is repeated for solutions of $\lambda_{MDMA}(u,L,H)=\lambda_{DFA}(u,L,H)$. They are drawn in Figs.\,10a,\,10c and in Figs.\,10b,\,10d respectively. Each of these contours divides $(u,L)$ plane into two regions: top right region above the corresponding curve where the scaling range for DFA  is larger than for DMA, and the bottom left area where it is on the opposite (see Figs.\,10a,\,10c). The similar graphic representation for $\lambda_{MDMA}$ vs $\lambda_{DFA}$ scaling ranges as a function of $L$, $u$ and $H$ is drawn in Figs.\,10b,\,10d. The absence of contours for MDMA for $H=0.7$ and $H=0.8$ in Fig.\,10b ($CL=97.5\%$) and for $H=0.6$, $H=0.7$, $H=0.8$ in Fig.\,10d ($CL=95\%$) indicates that for these autocorrelated series the scaling range for MDMA is always larger than the one for DFA . The top points terminating all contours show the maximal length of consecutive autocorrelated time series for which an intersection of scaling ranges between DFA and DMA or DFA and MDMA may occur (see also Figs.\,8,\,9). For longer series with the particular level of autocorrelations $\lambda_{DMA}$ (or $\lambda_{MDMA}$) always exceeds  $\lambda_{DFA}$ for any chosen $u$ value.

 All the revealed relations are believed to make an useful tool in determination of scaling ranges,
especially if there is a need to consider large data sets
arranged in a big number of shorter time subseries, e.g. in a search for evolving (time-dependent)
local Hurst exponent in various areas and applications.

 It is worth to discuss the efficiency of all three methods in a proper determination of $H$ scaling exponent which takes the precisely determined scaling ranges into account. It is beyond the scope of this article and will be a subject of forthcoming study \cite{in_preparation}.

\begin{figure}
\begin{center}
{\psfig{file=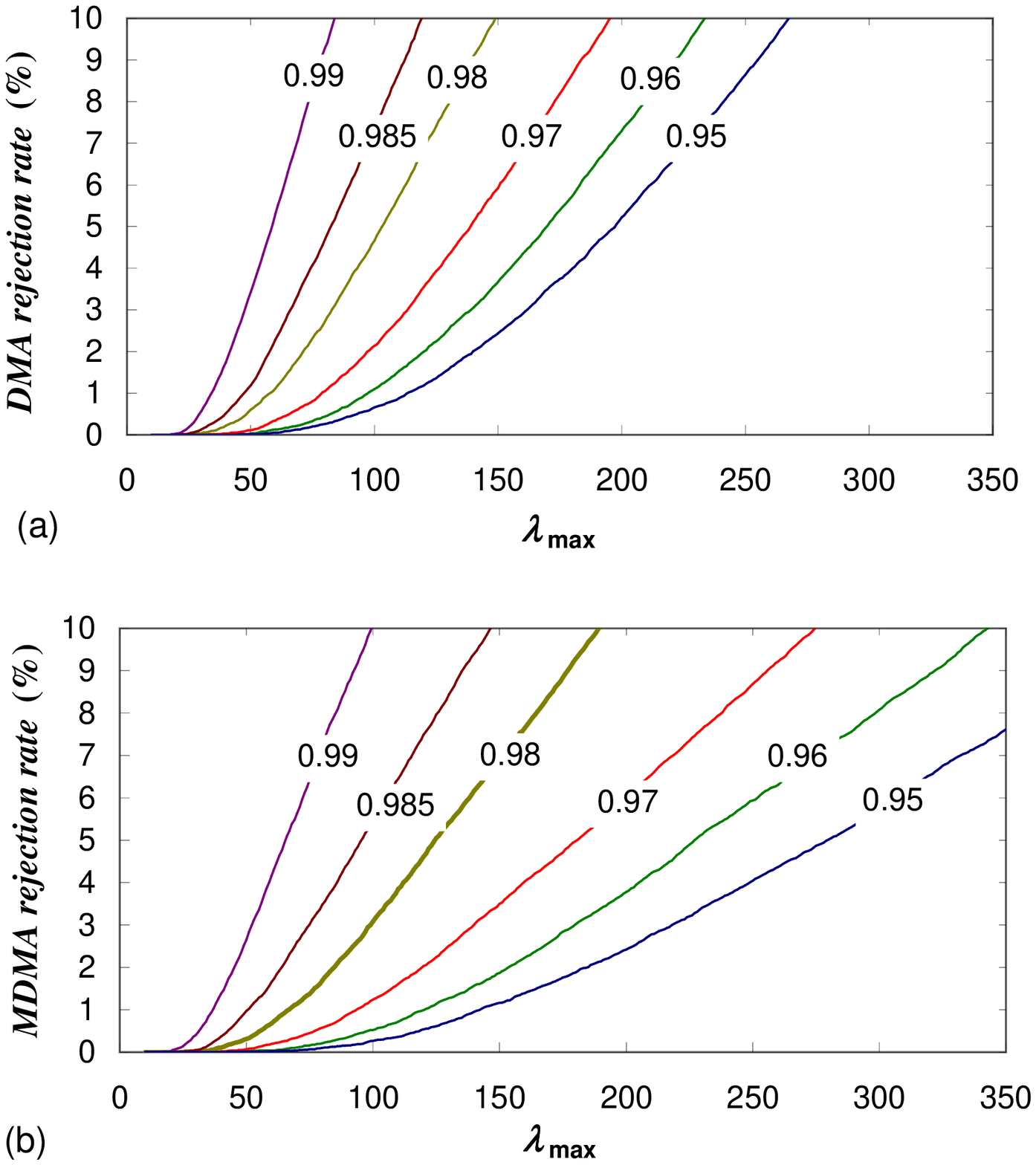,width=15cm}}
\end{center}
\caption{Percentage rate (\%) of rejected time series as a function of scaling range $\lambda_{max}$ and goodness of fit $R^2$ drawn for DMA (a) and MDMA (b)}
\end{figure}

\begin{figure}
\begin{center}
{\psfig{file=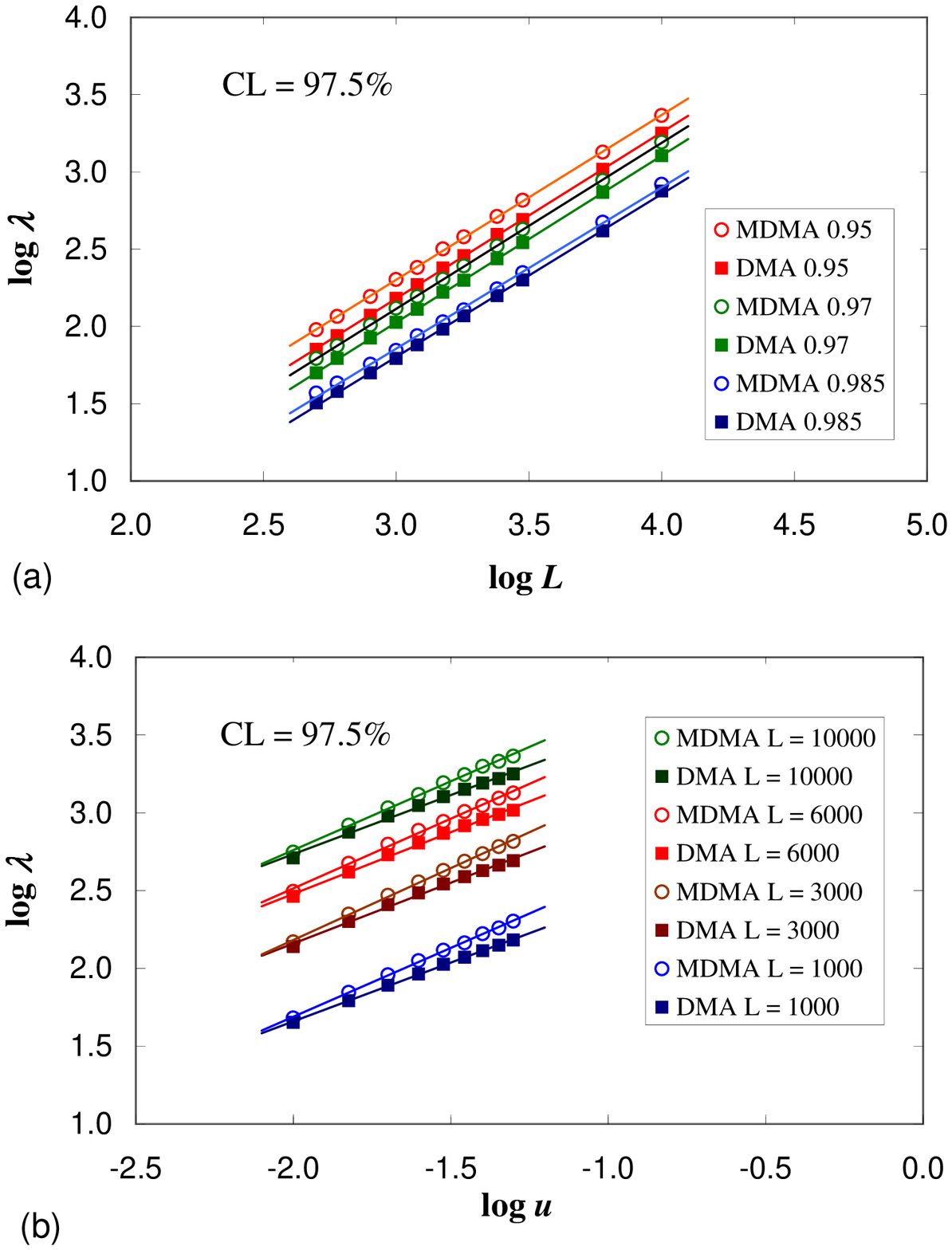,width=15cm}}
\end{center}
\caption{Dependence between scaling range $\lambda$ and total length of data $L$ (a) and goodness of fit $u=1-R^2$ (b) in log-log scale for DMA and MDMA at the confidence level $CL=97.5\%$}
\end{figure}

\begin{figure}
\begin{center}
{\psfig{file=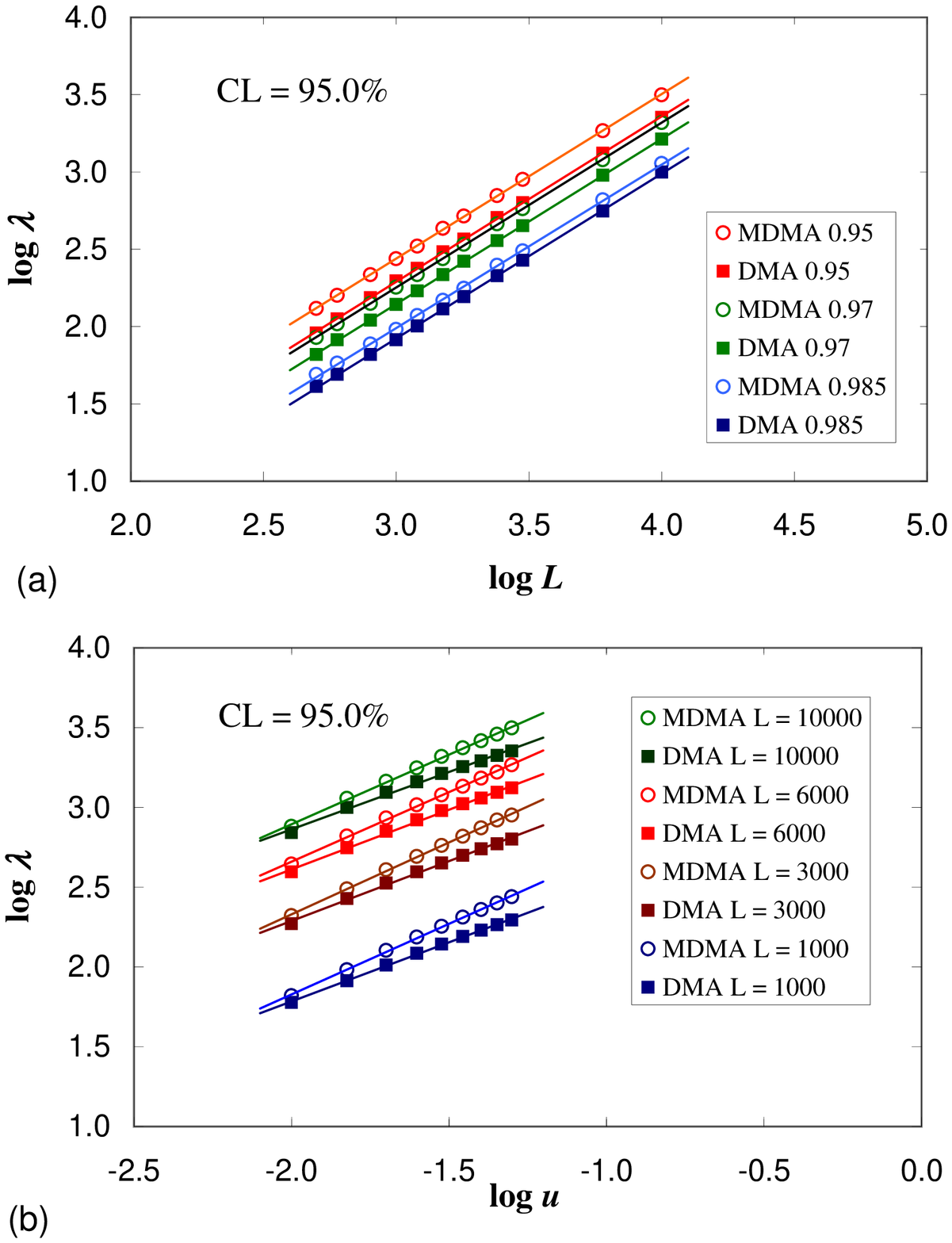,width=15cm}}
\end{center}
\caption{Same as in Fig.\,2a,\,2b but for $CL=95\%$}
\end{figure}

\begin{figure}
\begin{center}
{\psfig{file=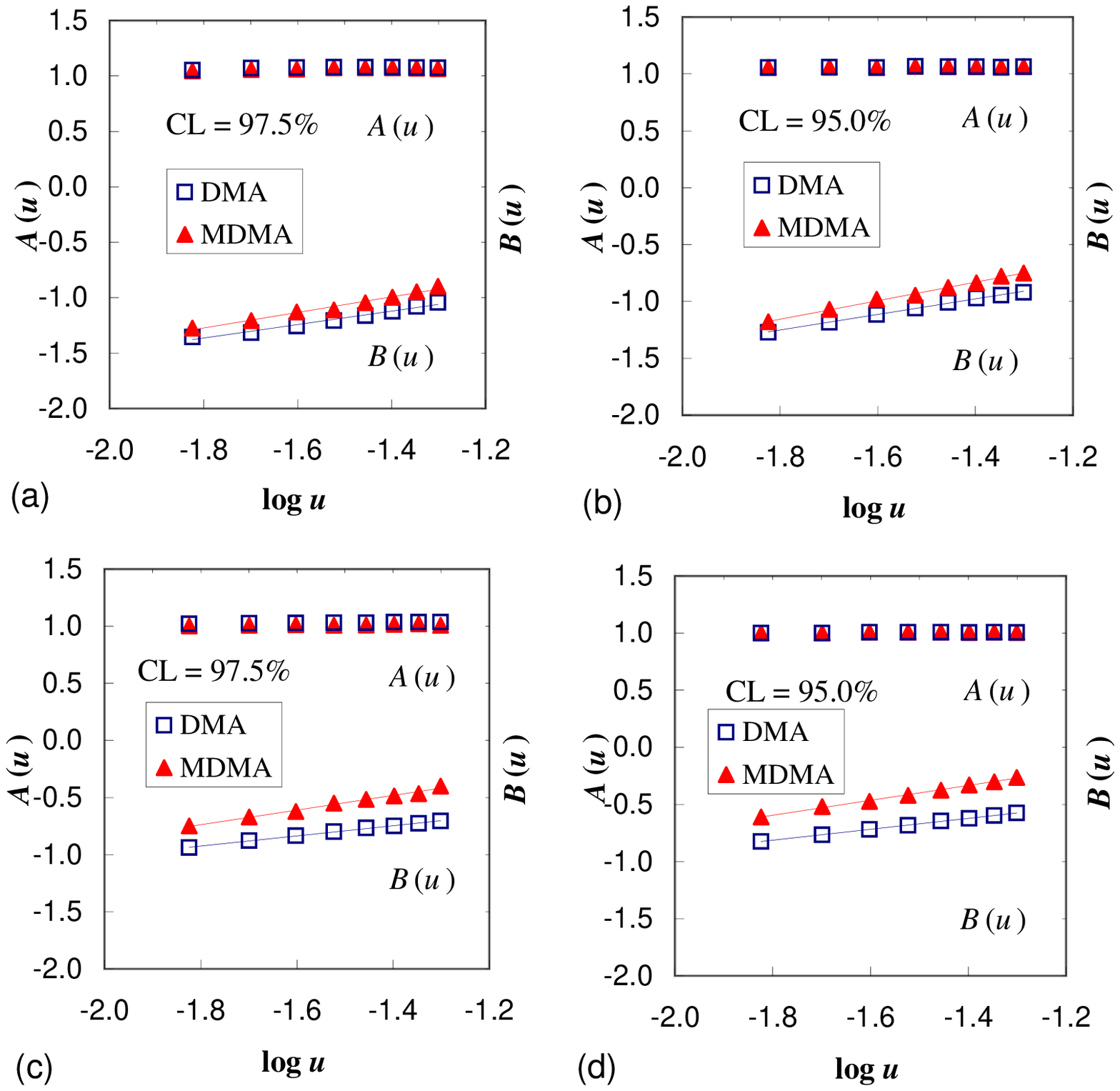,width=15cm}}
\end{center}
\caption{Dependence of $A$ and $B$ coefficients from Eq.(12) on $u=1-R^2$. It is seen that $A(u)$ does not depend on $u$. Top two panel (a), (b) correspond to uncorrelated series ($H=0.5$), while bottom ones (c),(d) are related to strongly autocorrelated data ($H=0.8$).}
\end{figure}

\begin{figure}
\begin{center}
{\psfig{file=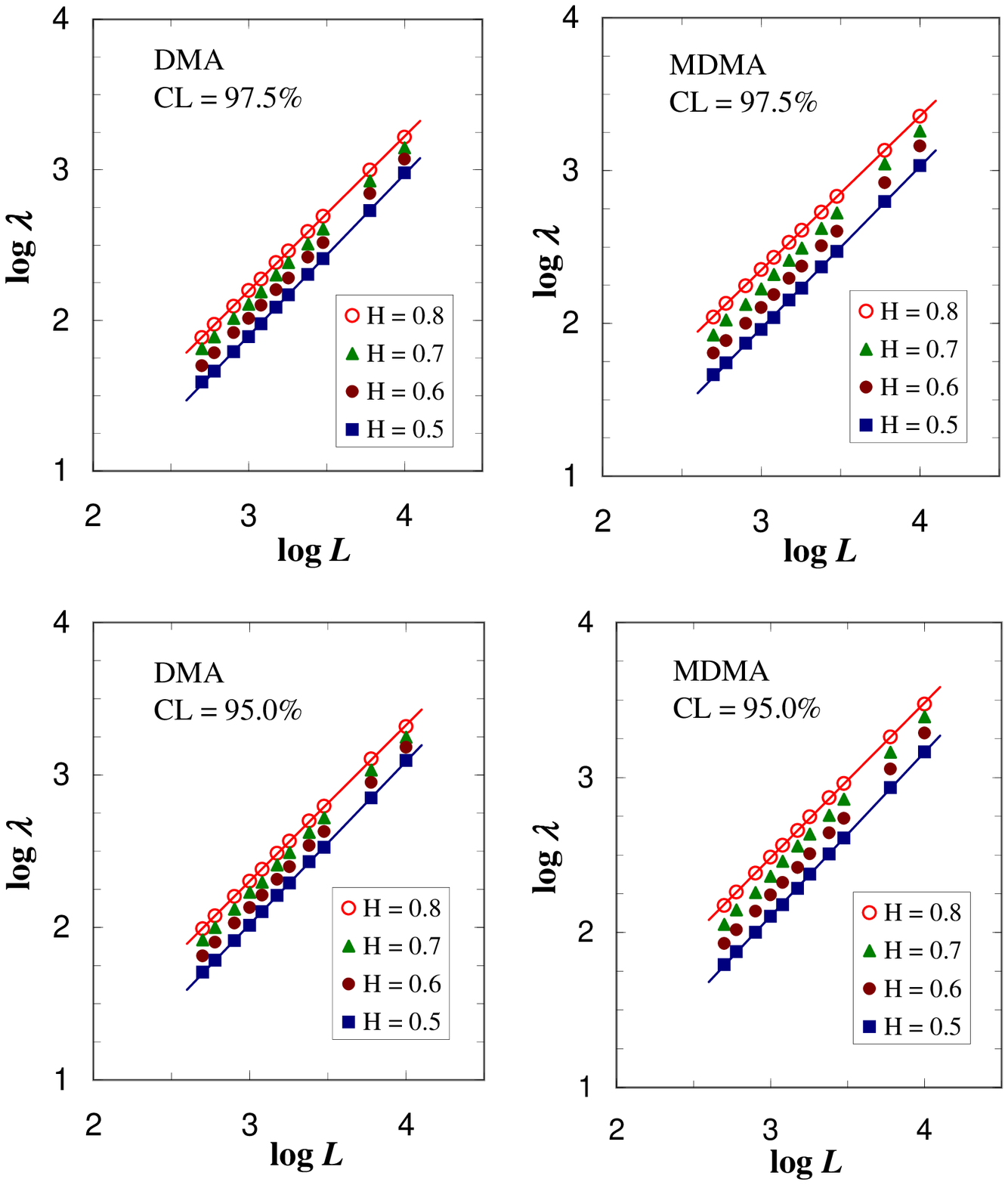,width=15cm}}
\end{center}
\caption{Dependence between scaling range $\lambda$ and length of time series $L$ for various levels of autocorrelation in data (measured by  $H$ exponent). Plots are drawn for particular choice $R^2= 0.98$ and look qualitatively the same for other $R^2$ values (not shown). Perfect linear dependence $\log\lambda(\log L)$ is observed in whole domain $0.5\leq H \leq 0.8$ for both DMA and MDMA methods. Fitting lines are shown only for the edge values $H=0.5$ and $H=0.8$ to make all remaining plots more readable.}
\end{figure}

\begin{figure}
\begin{center}
{\psfig{file=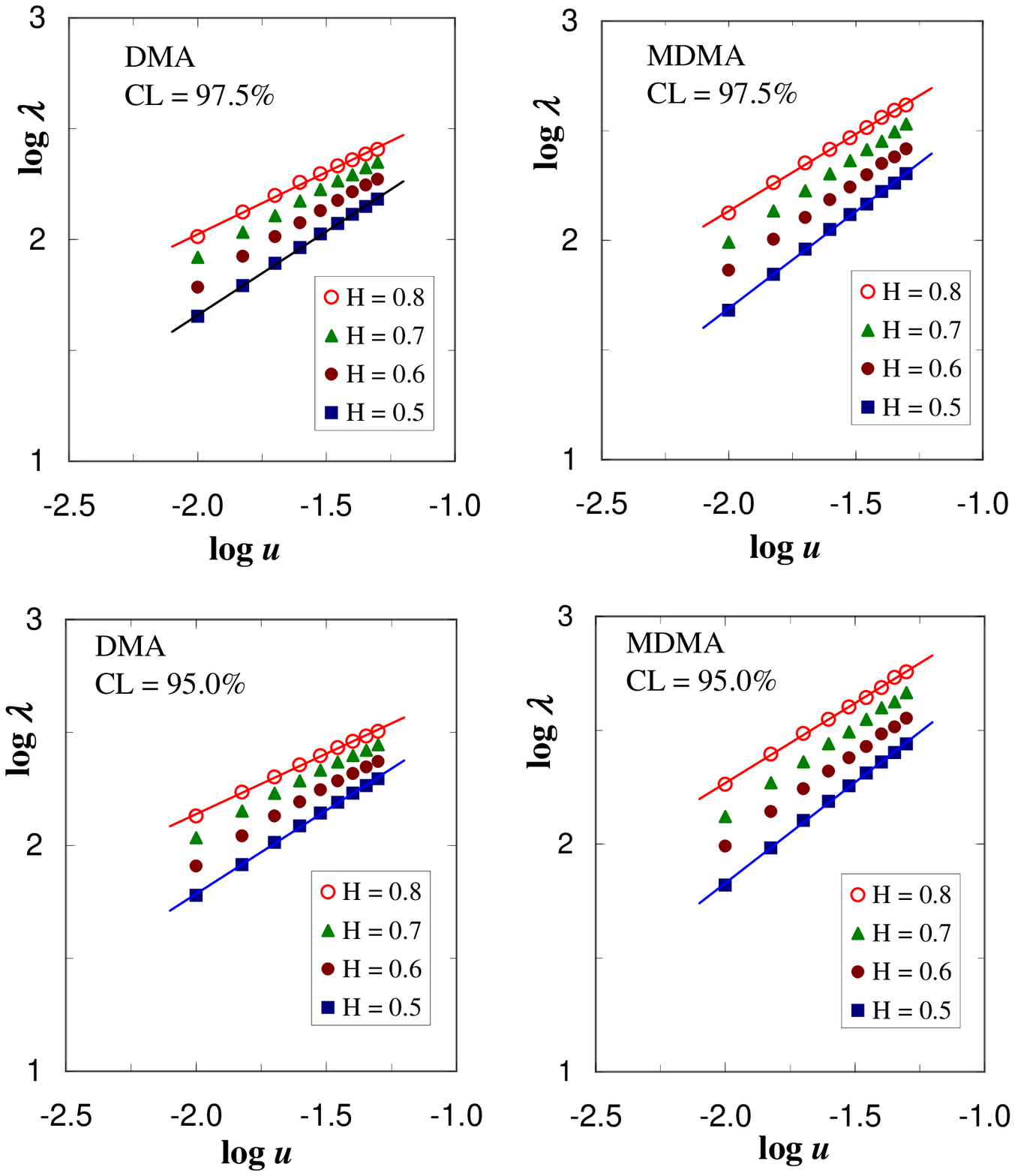,width=15cm}}
\end{center}
\caption{Same as in Fig.\,5 but for $\lambda$ dependence on $u=1-R^2$ with fixed length of data $L=10^3$. Results for other data lengths (not shown) are qualitatively identical.}
\end{figure}

\begin{figure}
\begin{center}
{\psfig{file=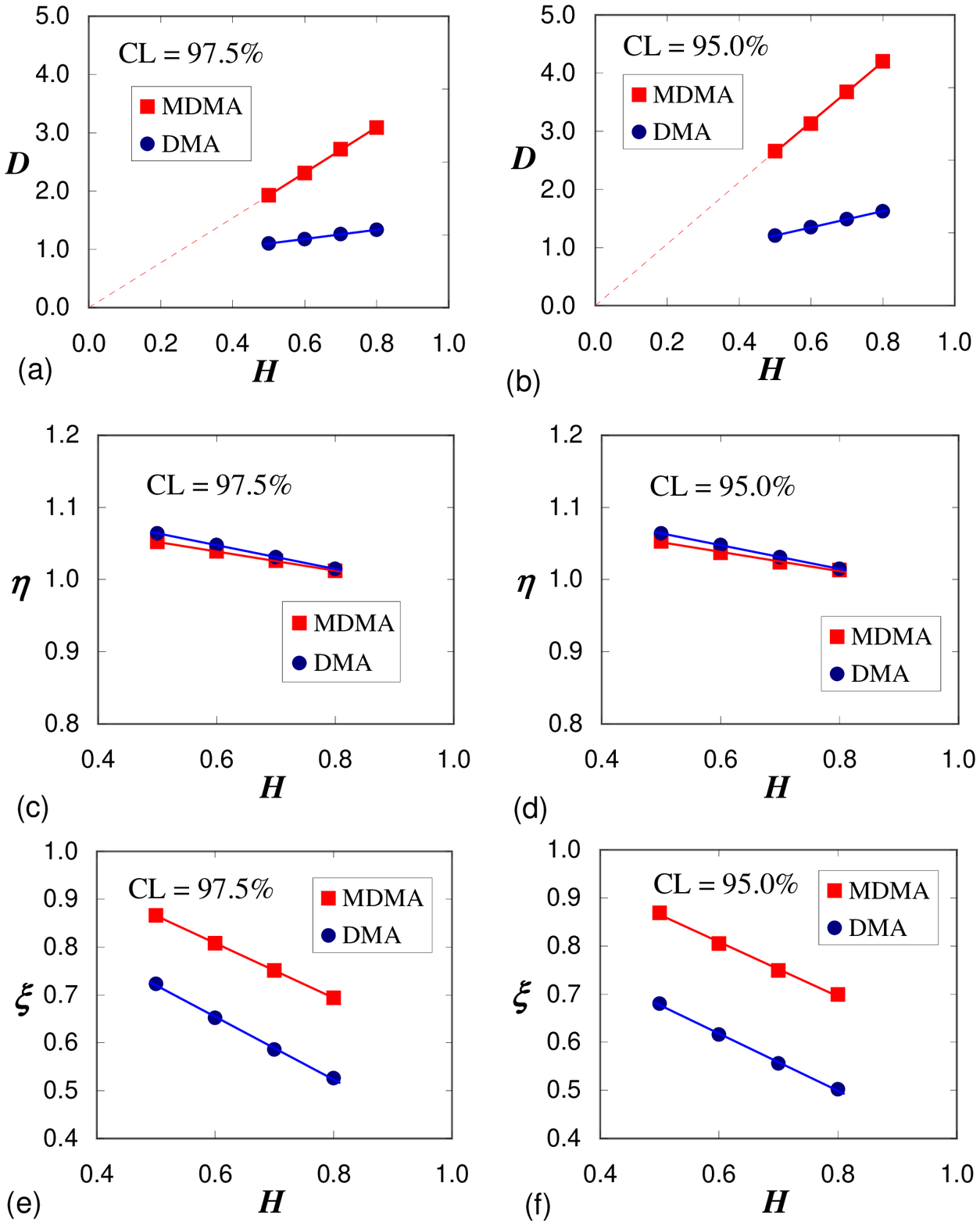,width=15cm}}
\end{center}
\caption{Dependence of the best fit parameters $D$, $\eta$, $\xi$ from Eq.(14) on the Hurst exponent $H$ (autocorrelation level) for DMA and MDMA methods for two confidence levels: $CL=97.5\%$ and $CL=95\%$. Dotted lines in two top panels, making an extension of the fit $D(H)$  towards $H=0$ for MDMA, prove that $D^{(2)}_0 = 0$ (see also Eq.(17)).}
\end{figure}

\begin{figure}
\begin{center}
{\psfig{file=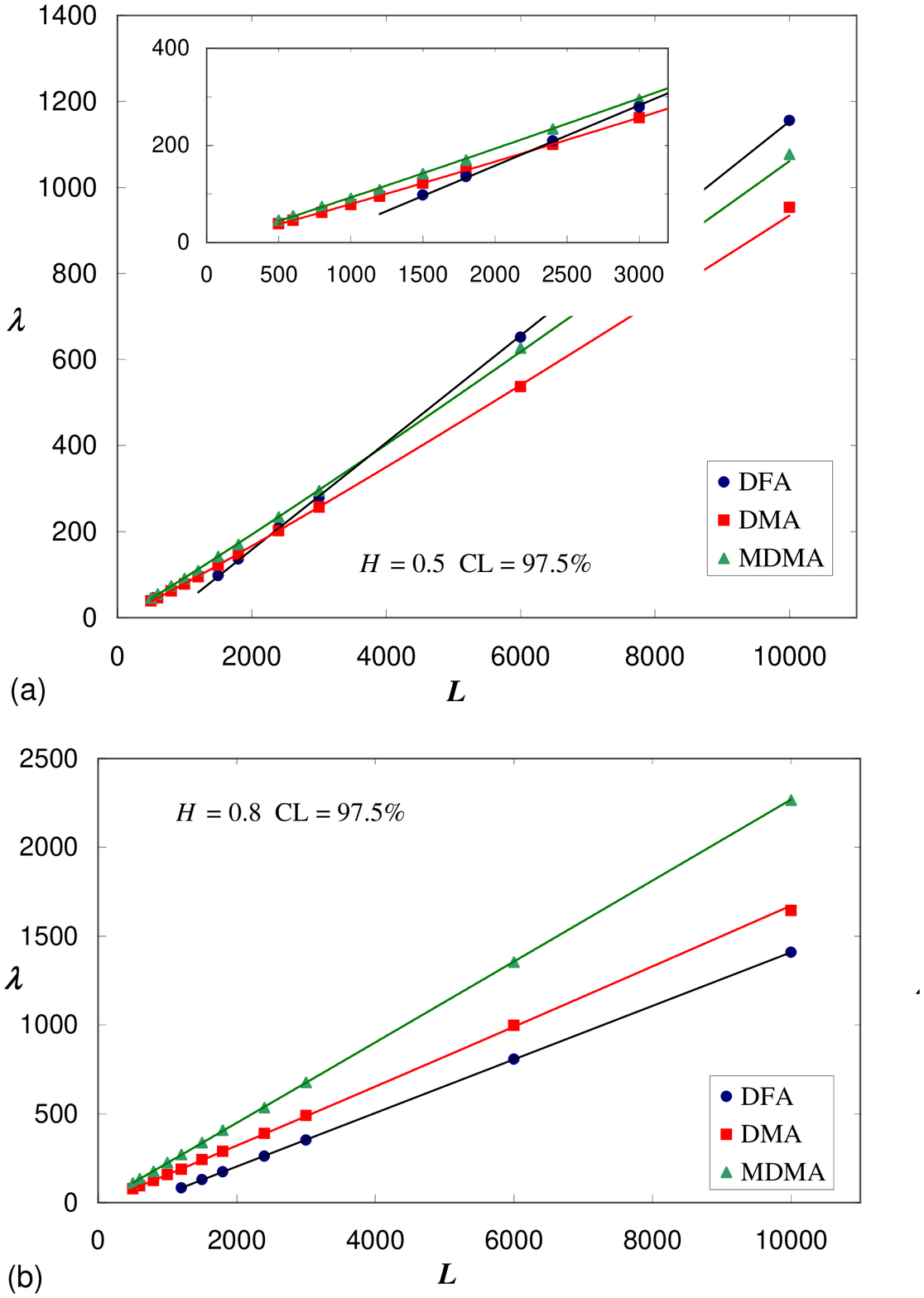,width=12.5cm}}
\end{center}
\caption{Hierarchy of scaling ranges for DFA, DMA and MDMA methods given as a function of series data length $L$ for uncorrelated (a) and strongly autocorrelated (b) signals for $CL=97.5\%$. Inbox in top panel shows details of this hierarchy for very short series ($L\leq 3\times 10^3$). It is evident how the hierarchy changes with increasing $H$ (compare (a) and (b) panels). Note that for uncorrelated data ($H=0.5$) $\lambda_{DFA}>\lambda_{MDMA}>\lambda_{DMA}$ if $L\gtrsim 3\times 10^3$, while  $\lambda_{MDMA}>\lambda_{DMA}>\lambda_{DFA}$ for arbitrary length of data for $H=0.8$.}

\end{figure}

\begin{figure}
\begin{center}
{\psfig{file=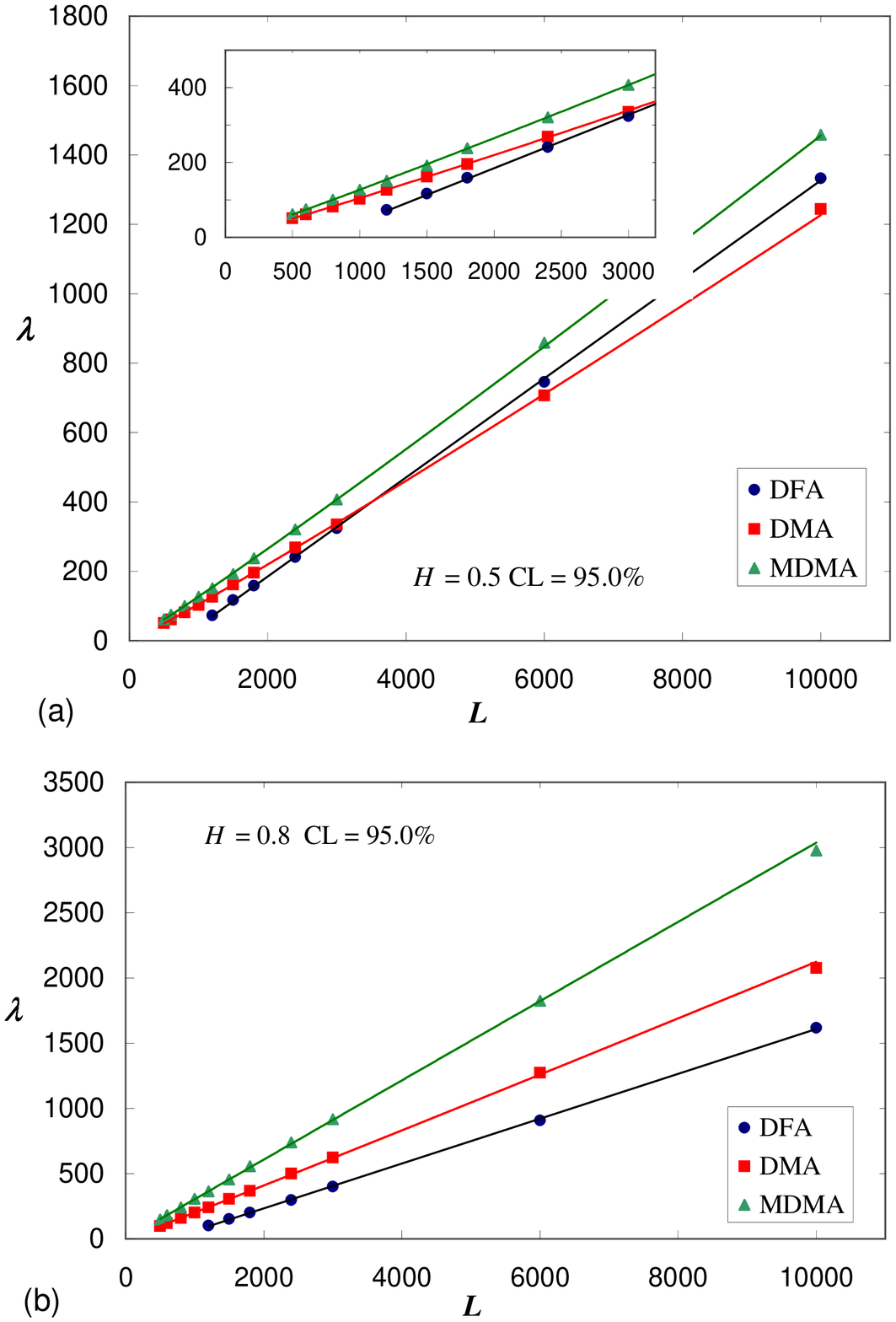,width=13cm}}
\end{center}
\caption{Same as in Fig.\,8 but for less restrictive confidence level $CL=95\%$. Here $\lambda_{MDMA}>\lambda_{DFA}>\lambda_{DMA}$ for uncorrelated data for $L\gtrsim 3\times 10^3$ (a) on the contrary to Fig.\,8a.}
\end{figure}

\begin{figure}
\begin{center}
{\psfig{file=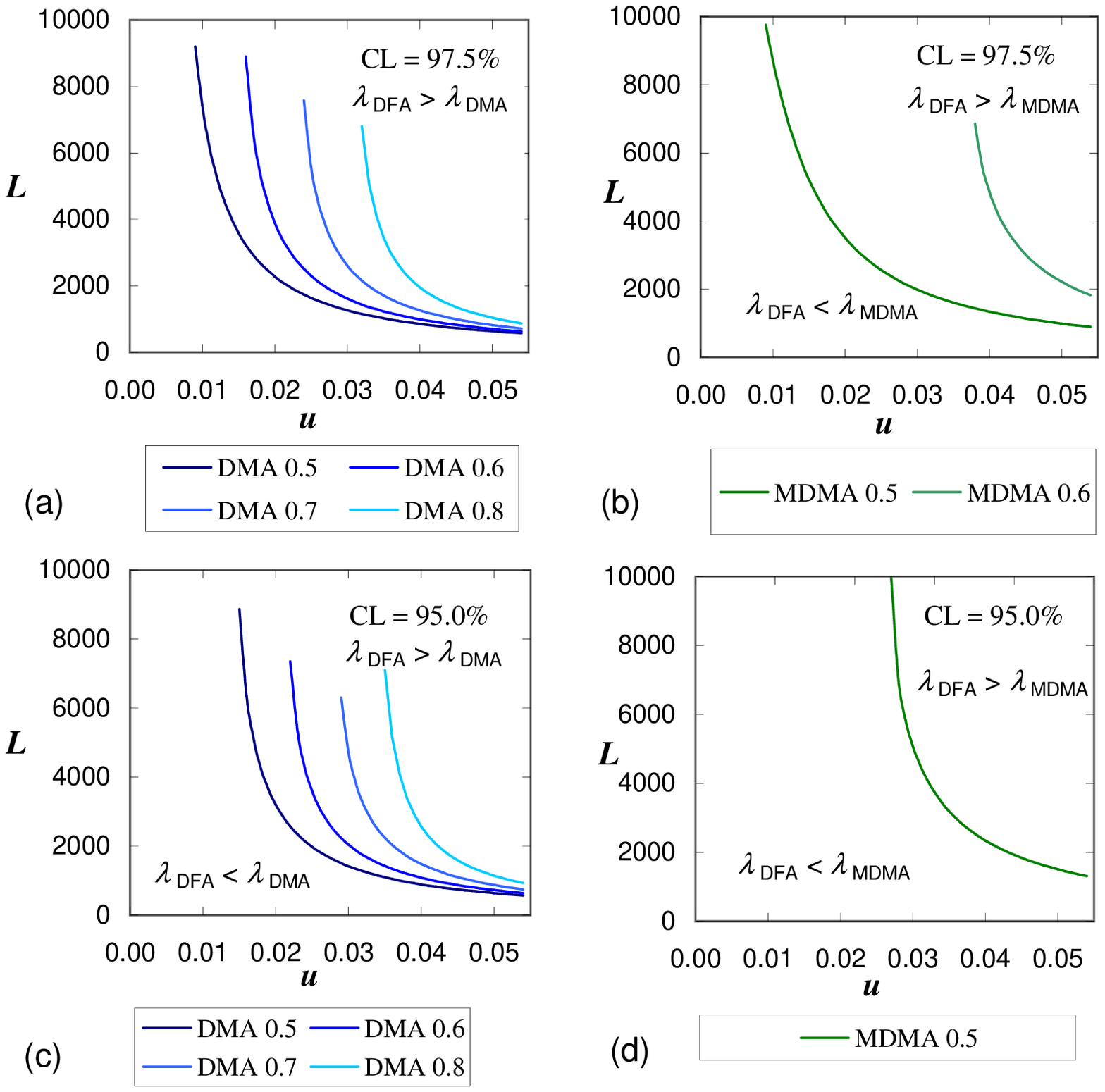,width=15cm}}
\end{center}
\caption{Solutions in $(u,L)$ plane of the relationships: $\lambda_{DMA}(u,L,H)=\lambda_{DFA}(u,L,H)$ (a,c), $\lambda_{MDMA}(u,L,H)=\lambda_{DFA}(u,L,H)$ (b,d) for fixed values of Hurst exponent $H$ and $CL$. Top right part of $(u,L)$ area above each contour corresponds to $(u,L)$ values where $\lambda_{DFA}>\lambda_{DMA}$ (Fig.\,10a,\,10c) or  $\lambda_{DFA}>\lambda_{MDMA}$ (Fig.\,10b,\,10d). Bottom left part of all figures indicates the opposite situation.}
\end{figure}

\end{document}